\documentclass[%
reprint,
showpacs,preprintnumbers,
 amsmath,amssymb,
 aip,
 graphicx,
apl,
]{revtex4-1}

\usepackage{graphicx}% Include figure files
\usepackage{dcolumn}% Align table columns on decimal point
\usepackage{bm}% bold math
\usepackage[mathlines]{lineno}% Enable numbering of text and display math
\begin{document}
\preprint{} 

\title[Short Title]{Coherent monochromatic phonons in highly-purified semiconducting single-wall carbon nanotubes}

\author{Yuki Honda$^{*}$}
\author{Elizabeth Maret$^{*}$}
\affiliation{Institute of Applied Physics, University of Tsukuba, 1-1-1 Tennodai, Tsukuba 305-8573, Japan}
\author{Atsushi Hirano}
\author{Takeshi Tanaka}
\affiliation{Nanosystem Research Institute, National Institute of Advanced Industrial Science and Technology, Tsukuba Central 4, 1-1-1 Higashi, Tsukuba 305-8562, Japan}
\author{Kotaro Makino}
\author{Muneaki Hase}
\email{mhase@bk.tsukuba.ac.jp}
\affiliation{Institute of Applied Physics, University of Tsukuba, 1-1-1 Tennodai, Tsukuba 305-8573, Japan}

\date{\today}

\begin{abstract}
We have used a femtosecond pump-probe impulsive Raman technique to explore the polarization dependence of coherent optical phonons in highly-purified and aligned semiconducting single-wall carbon nanotubes (SWCNTs). Coherent phonon spectra for the radial breathing modes (RBMs) exhibit a different monochromatic frequency between the film and solution samples, indicating the presence of differing exciton excitation processes. By varying the polarization of the incident pump beam on the aligned SWCNT film, we found that the anisotropy of the coherent RBM excitation depends on the laser wavelength, which we consider to be associated with the resonant and off-resonant behavior of RBM excitation.  
\end{abstract}

\pacs{78.47.J-, 63.22.Np, 63.20.kd,}

\maketitle
Single-wall carbon nanotubes (SWCNTs) are the simplest forms of carbon nanotubes, existing as essentially 1-D systems formed by folding and conjoining two ends of 2-D graphene sheets.  SWCNTs display exceptional mechanical, electrical and optical properties.  One area of particular interest of SWCNT characteristics is in their varying electrical properties, either metallic ($m$-SWCNT) or semiconducting ($s$-SWCNT), based on their chirality\cite{Jiang07}.  SWCNTs are expected to have tremendous applications over a broad range of fields including chemical sensors\cite{Modi03} and nano-machines applied to medical science\cite{Zhang08}.  Another interest, which we attempted to explore in this study, lies in discerning the optical phonon modes of SWCNT, because of their potential for optical manipulation of structures\cite{Traian04,Makino11}.
 
Coherent phonon spectroscopy\cite{Dekorsy00}, stimulated and evaluated by ultrafast pump-probe technique, can be used to study SWCNT carrier dynamics and evaluate their diameter size\cite{Hertel00}.  Samples of SWCNTs are often obtained as mixtures of both $m$-SWCNTs and $s$-SWCNTs.  
In such mixtures of SWCNT, the complexity of different electronic band structures as well as the distribution of nanotube diameters make the optical response a multi-peak function, as has been observed by Raman spectroscopy\cite{Pimenta98,Saito03}.  
Coherent phonon spectroscopy can thus be used to evaluate the spectrum corresponding to the distribution of the diameters in a sample as well as elicit concurrent expansion and contraction along the length of the SWCNT, exciting the radial breathing mode (RBM)\cite{Hertel05,Lim06,Gambetta06,Kato08,Kim09,Luer09}.     
 
It is well known that coherent phonon spectroscopy exhibits a large anisotropic phonon response dependent upon the polarization of excitation light  \cite{Hwang00, Kim09, Booshehri11}. However, little is known about the behavior of pure (metal/semiconductor separated) SWCNT samples excited at resonant and off-resonant wavelengths.  In this paper, we present the investigation of the anisotropic properties of two pure and separated $s$-SWCNT samples, one as an aligned film and the other as a dispersed solution. The samples were excited and probed under resonant (810 nm) and off-resonant (850 nm) conditions (Fig. 1) using coherent phonon spectroscopy.  Our current results reveal that there are spectral differences between the $s$-SWCNT solution and film samples in the coherent RBM due to the different resonant conditions for excitons. 
The coherent phonon dynamics also exhibit the effect of resonance on the polarization dependence of the RBM intensity such that the polarization anisotropy is more intense under resonant excitation at 810 nm. To the best of our knowledge, this behavior has not been previously observed in pure SWCNT samples.
 \begin{figure}[htbp]
\includegraphics[width=80mm]{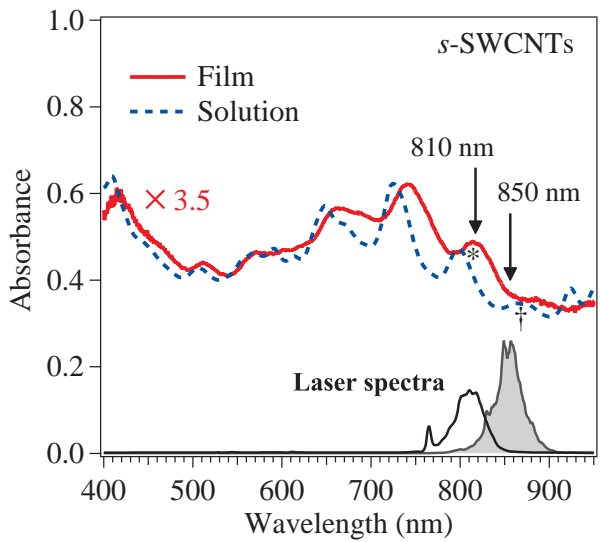}
\caption {(color online) Absorption spectra of the $s$-SWCNT film and solution in the visible to near-infrared region. The spectra of the laser (peaks at 850 nm and 810 nm) are also shown with gray and white colors  to understand which excitonic resonance is excited by the laser in each case; * represents the resonance in the film and $\dagger$ is that in solution when the central wavelength is 850 nm.}
\label{Fig. 1}
\end{figure}

The samples of SWCNTs were produced via a high-pressure catalytic CO decomposition (HiPco) process.  Sodium dodecyl sulfate (SDS) was used to disperse SWCNTs in an aqueous solution.  The SWCNTs were then dispersed by ultrasonication for 1 hour, yielding a SWCNT density of 1 mg/ml. 
After the ultracentrifugation at 210,000 $\times$ g for 1 hour, upper 70\% of the supernatant was collected and then poured on a column containing gel (Sephacryl S200) to separate the SWCNTs into $m$-SWCNTs and $s$-SWCNTs \cite{Hirano12}.

The aligned $s$-SWCNT films were prepared by immersing a glass sample slide vertically into the separated $s$-SWCNT solution \cite{Shimoda02}. Fig. 1 shows the absorption spectra of $s$-SWCNT for the film and solution samples. The characteristic bands at 580 - 900 nm are assigned to the S$_{22}$ transition. Peaks corresponding to the exciton bands\cite{Wang05} and background caused by $\pi$-plasmon resonant absorption are clearly visible\cite{Landi05}. The spectrum of the film exhibits red-shifted and broader exciton bands, compared to the solution, which can possibly be explained by the effects of Van der Waals interactions or dependence of the dielectric function due to the surroundings \cite{Perebeinos04,Fantini04,Sheng05}.

Time-resolved transient transmission $\Delta T/T$ and reflection $\Delta R/R$ of the sample were measured by employing a fast scanning delay technique\cite{Cho90,Hase03} using a mode-locked Ti:sapphire laser. Nearly collinear pump and probe pulses (20 fs pulse duration; 80 MHz repetition rate) were focused to a 70 $ \mu $m spot onto the sample\cite{Makino09}. The average power of the pump and probe beams were fixed at 40 and 3 mW for the transmission and reflection measurements, respectively. The transient transmission and reflection were recorded as functions of time delay $\tau$ between the pump and probe pulses. 
We utilized an optical pump-probe technique with two different excitation wavelengths, 810 nm (strongly resonant for the SWCNT film) and 850 nm (off-resonant for SWCNT film, but near-resonant with SWCNT solution). Thus, we were able to investigate the effects of resonant and off-resonant excitation on the dynamics of coherent RBMs in highly purified $s$-SWCNTs. The polarization angle of the pump beam ($\theta$) was varied from 0$^\circ $ to 90$^\circ $ by manually rotating the half-wave plate, while that of the probe beam was kept at 0$^\circ $, where $\theta$ = 0$^\circ $ means the polarization is parallel to the SWCNT alignment and $\theta$ = 90$^\circ$ corresponds to the polarization is perpendicular to the SWCNT alignment. 

\begin{figure}[htbp]
\includegraphics[width=80mm]{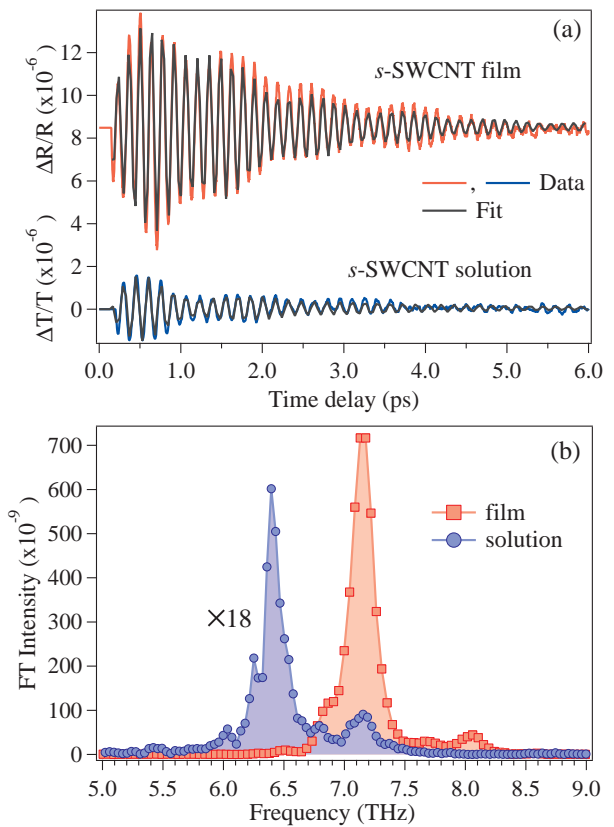}
\caption {(color online) (a) Time-resolved $\Delta R/R$ and $\Delta T/T$ signals observed in $s$-SWCNT film and solutions at 850 nm. The black solid lines represent the fit with damped harmonic oscillations. (b) FT spectra obtained from the time domain data in (a). The FT intensity for the solution is magnified by a factor of 18.}
\label{Fig. 2}
\end{figure}
Figure 2(a) shows the time-resolved $\Delta R/R$ and $\Delta T/T$ signals observed in the $s$-SWCNT film and solutions, respectively. The laser wavelength used was 850 nm. 
The coherent RBM oscillations were observed in both samples with a much larger amplitude for the case of the SWCNT film. The RBM oscillations were well fit to a linear combination of three damped harmonic oscillations (Fig. 2(a) black line). One aspect to notice from the obtained parameters is that the relaxation time of the dominant coherent RBM oscillation in the film (1.6 ps) is significantly shorter than that of the solution (2.2 ps), which can be interpreted as being associated with a larger phonon damping in the aligned films due to the stacked structure of the aligned film\cite{Shimoda02}. Also the data shows that the coherent RBM in the film exhibits less beating between multiple modes than those of the un-purified solutions\cite{Lim06, Kim09, Makino09}, which is explained by the high purity of the $s$-SWCNT sample. 

In order to investigate the dynamics of the coherent RBM, the time-domain data in Fig. 2(a) were converted into FT spectra, as displayed in Fig. 2(b). The FT spectra indicate that the main peak appears at 7.2 THz for the film sample while for the solution sample the response intensity drops by an order of magnitude and the peak appears at 6.4 THz. The observation of the monochromatic coherent RBM in the present study for both the film and solution verifies the high purity of our sample, so that only a single chirality RBM is excited. 
The frequency change (0.8 THz) of these main peaks can be understood by the difference in the absorption peaks corresponding to the resonant excitons as shown in Fig. 1. 
Considering the laser spectrum used in the present study (Fig. 1), the photo-excitation should occur dominantly at 810 nm exciton peak in the SWCNT film, while the excitation should occur at 860 nm exciton peak in the SWCNT solution. Thus the difference in the exciton resonance is considered to be a result of the difference in the diameter $R_{t}$ of the excited SWCNT. 
In fact, the frequency of the RBM ($\omega_{RBM}$) estimated from the relation between the photon energy\cite{note1} and the tube diameter $R_{t}$ in the Kataura's plot\cite{Kataura99} and the relation of $\omega_{RBM}$ = 7.44 [THz]/$R_{t}$ [nm] \cite{Makino09}, 
is 7.22 THz for the film and 6.47 THz for the solution, which are in good agreement with the experimental results in Fig. 2(b). Note that the change in tube diameter by pressure effect due to the bundle structure (film) cannot account for the 0.8 THz difference of the RBM frequency because the RBM shift has been found to be only $\sim$ 0.2 THz even at 1 GPa pressure\cite{Lebedkin06}. 

\begin{figure}[htbp]
\includegraphics[width=80mm]{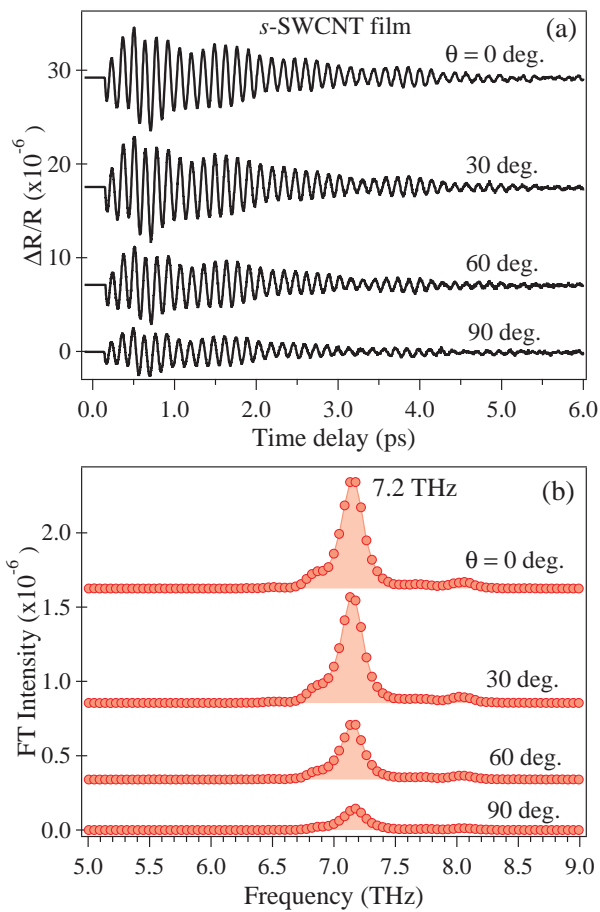}
\caption {(color online) (a) Comparison of the time-resolved $\Delta R/R$ signals observed in the film of $s$-SWCNT at different polarization angles. (b) Corresponding FT spectra obtained from the time-domain data in (a). }
\label{Fig. 3}
\end{figure}
Figure 3 compares the coherent RBM oscillations observed in the aligned film excited by 850 nm light at different pump polarization angles, indicating the polarization dependence of the RBM amplitude, as has been observed in a similar experiment \cite{Booshehri11}. The FT spectra of the coherent RBM oscillations were obtained as shown in Fig. 3 and the integrated spectral intensity was plotted in Fig. 4 for the two different excitation wavelengths (810 nm and 850 nm) after normalization of the integral. The data were well fit to the phenomenological function $Acos^{p}(\theta)$, where $A$ is the intensity, $p$ is the fitting parameter, and $\theta$ is the polarization angle.  Considering $p$ is determined by the slope of the integrated FT spectra data, it indicates the degree of anisotropic dependence.  Thus a larger $p$ value, and hence a steeper slope, indicates a greater anisotropic dependence, while a lower $p$ value indicates reduced anisotropic dependence. In the case of off-resonant excitation at 850 nm, the value of the fitting parameter $p$ is  2.1$\pm$ 0.1, while $p$ is 3.7$\pm$ 0.3 for near-resonant excitation at 810 nm \cite{note2}. 
Booshehri {\it et al}. reported that $p$ is $\approx$ 3.6 in the case of the resonant excitation for a film sample made of the mixture of $m$-SWCNTs and $s$-SWCNTs, although it must be noted that sample preparation and hence the actual film structure was significantly different for ours. 
Booshehri {\it et al} further predicted the $p$ value based on a theoretical model and found that $p$ $\approx$ 4 for the resonant excitation with $E_{44}$ transition \cite{Booshehri11}. 
A plausible reason why the value of $p$ is $\sim$ 2 for the case of off-resonant excitation is that there is less contribution from the exciton absorption and dominant contribution from the free carrier absorption due to $\pi$-plasmon interactions provide the dominant response; the latter has less anisotropy than the former. 
Note that the value of $p$ is $\approx$ 2.2 - 2.4 in the case of the $s$-SWCNT solution, implying a $A[1 + cos^{2}(\theta)]$ behavior as demonstrated in SWCNT solutions\cite{Kim09}. 
\begin{figure}[htbp]
\includegraphics[width=70mm]{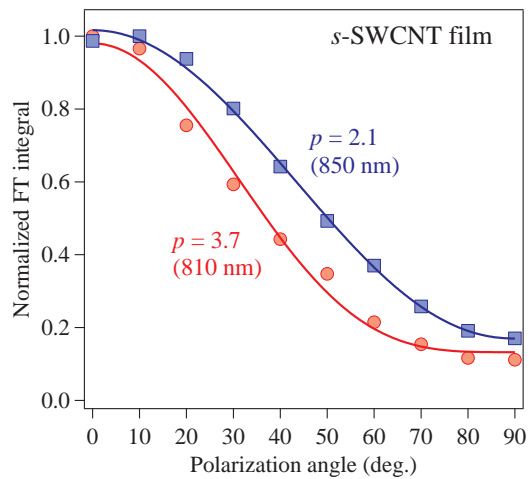}
\caption {(color online) FT integral of the RBM peak at 7.2 THz for the two different wavelengths as a function of pump polarization. }
\label{Fig. 4}
\end{figure}

In summary, we have explored the ultrafast dynamics of coherent RBM in real time using the femtosecond pump-probe technique in well-aligned SWCNT film. Coherent phonon spectra for the RBM exhibit the different monochromatic frequency between the film and solution samples because of the presence of differing exciton resonances. By varying the incident pump polarization on the aligned SWCNT film, we found that the anisotropy of the coherent RBM excitation depends on the laser wavelength, which we consider to be associated with the resonant and off-resonant behavior of RBM excitation.  

%Acknowledgements%
E. M. acknowledges the support of the Fulbright Fellows Grant from Fulbright Program, U.S. Department of State. 
This work was supported in part by KAKENHI-23104502 under the Scientific Research on Innovative Areas, Materials Design through Computics: Complex Correlation and Non-equilibrium Dynamics from MEXT, Japan. 

$^{*}$These authors contributed equally to this work.

\pagebreak


\begin{thebibliography}{99}
\bibitem{note2}
We have measured the coherent RBM by exciting and probing also at 790 nm and 890 nm (off-resonant) and the values of $p$ are obtained to be $\approx$ 2.0.
\end{thebibliography}
\end{document}